\documentclass[%
 reprint,
superscriptaddress,
 amsmath,amssymb,
 aps,
]{revtex4-2}

\usepackage{graphicx}
\usepackage{dcolumn}
\usepackage{bm}
\usepackage[colorlinks=true, citecolor=blue, linkcolor=blue, urlcolor=blue]{hyperref} 
\usepackage{float}
\usepackage{placeins}

\begin{document}

\preprint{APS/123-QED}

\title{Optimal neutralization of negative space charges in photon-enhanced thermionic emission devices under bidirectional discharge}

\author{Xinqiao Lin}
\affiliation{Department of Physics, Xiamen University, Xiamen 361005, People’s Republic of China}
\author{Zhiqiang Fan}
\affiliation{Department of Physics, Xiamen University, Xiamen 361005, People’s Republic of China}
\author{Shunjie Zhang}
\affiliation{Department of Physics, Xiamen University, Xiamen 361005, People’s Republic of China}
\author{Xiaohang Chen}
\affiliation{Department of Physics, Xiamen University, Xiamen 361005, People’s Republic of China}
\author{Zhimin Yang}
\affiliation{School of Physics and Electronic Information, Yan’an University, Yan’an 716000, People’s Republic of China}
\author{Jincan Chen}
\affiliation{Department of Physics, Xiamen University, Xiamen 361005, People’s Republic of China}
\author{Shanhe Su}
 \email{sushanhe@xmu.edu.cn}
\affiliation{Department of Physics, Xiamen University, Xiamen 361005, People’s Republic of China}


\date{February 24, 2025}

\begin{abstract}

In this study, we innovatively modeled photon-enhanced thermionic emission (PETE) devices, incorporating positive ion injection and bidirectional discharge's effects on the space charge barrier simultaneously. Compared to previous models, our model allows the positive ion distribution function to be compatible with scenarios in which the anode motive is either higher or lower than the cathode motive, and also adapts to significant anode discharge. Through numerical simulations and parametric analyses, we found that: (1) As the ratio of the positive ion increases, the capability for space charge neutralization becomes stronger. (2) The lower the electron affinity is, the smaller the ratio of positive ions are required. (3) When the anode temperature is higher or the anode work function is lower, the impact of reverse discharge on the net current density is more pronounced. Conversely, when the anode temperature is higher or the anode work function is greater, the ratio of positive ions required to achieve complete space charge neutralization increases. This study further elucidates the mechanisms and characteristics of space charge neutralization effects in PETE devices, providing a theoretical foundation for optimizing their design. Additionally, the accompanying theory and algorithm possess the potential to spark innovative research across diverse fields.

\end{abstract}

\maketitle


\section{\label{sec:Introduction}Introduction}

Solar energy, as an inexhaustible and environmentally friendly resource, plays a crucial role in advancing carbon neutrality. Over the decades, research has significantly contributed to the development of high-efficiency solar cells\cite{Zhang_2022_AssessingEnergyTransition, Guo_2024_QuantumInterferenceRecombination, Li_2024_FlexibleSiliconSolar}. The thermionic emission converters (TECs), which directly convert a part of heat into electricity, have also seen significant advancements\cite{Li_2021_StudiesVirtualCathode, Kottke_2024_ImprovementThermionicEmission}, with efficiencies exceeding 30\%, as predicted by the Richardson-Dushman law\cite{Dushman_1923_ElectronEmissionMetals, Campbell_2021_ProgressHighPower}. Photon-enhanced thermionic emission (PETE) is a process that utilizes photons to enhance the efficiency of thermionic emission. By integrating elements of both photonic and thermal energy conversion, PETE emerges as a promising approach for solar energy harvesting\cite{Schwede_2010_PhotonenhancedThermionicEmission}. In recent years, PETE has garnered significant attention and has been extensively studied\cite{Diao_2024_TheoreticalExplorationPhotoemission, Li_2024_ImprovedMathematicalModel, Lin_2024_EffectSpaceCharge, Wang_2024_SolarCO2Splitting, Wang_2024_EffectSpaceCharge}. PETE devices demonstrate theoretical efficiencies exceeding 40\%, with the potential to reach 50\% when integrated with a heat engine.  However, a critical challenge for PETE devices is the space charge effect between the electrodes\cite{Su_2014_SpaceChargeEffects, Wang_2019_OptimalDesignInterelectrode, Rahman_2021_SemiconductorThermionicsNext, Lin_2024_EffectSpaceCharge}. This effect leads to the accumulation of electrons in the electrode gap, creating a barrier that suppresses cathode emission current, thereby resulting in a significant reduction in efficiency\cite{Hatsopoulos_1973_ThermionicEnergyConversion, Hatsopoulos_1979_ThermionicEnergyConversion, Smith_2007_ConsiderationsHighperformanceThermionic, Lee_2012_OptimalEmittercollectorGap, Smith_2013_IncreasingEfficiencyThermionic}. 

Numerous studies suggest that reducing the electrode gap is a theoretically promising strategy for mitigating space charge effects in PETE devices\cite{Su_2014_SpaceChargeEffects, Segev_2015_NegativeSpaceCharge, Lin_2024_EffectSpaceCharge}. Research has shown that when the electrode gap exceeds 5 micrometers, the energy conversion efficiency of thermionic devices can drop to as low as 20\% of their ideal value\cite{Lee_2012_OptimalEmittercollectorGap}. However, this approach faces two major challenges: substantial heat loss caused by near-field thermal radiation effects, as highlighted in\cite{Wang_2019_OptimalDesignInterelectrode, Rahman_2021_SemiconductorThermionicsNext, Qiu_2022_PhotothermoelectricModelingPhotonenhanced}, and the necessity for micro-scale manufacturing precision in metrology. Additionally, when the scale is reduced to the nano-scale, both quantum tunneling effects and image force corrections must be carefully considered\cite{Wang_2016_EffectsNanoscaleVacuum, Wang_2023_PerformanceAnalysisPhotonenhanced}. 

Fortunately, positive ion neutralization offers an alternative strategy to effectively suppress the space charge effect\cite{Mladenov_2001_PotentialDistributionSpacecharge, Ito_2012_OpticallyPumpedCesium, Wang_2024_EffectSpaceCharge}. Cesium atoms possess the lowest direct ionization potential of all metallic elements, with a measured value of just 3.89 eV\cite{Rasor_1991_ThermionicEnergyConversion}. In traditional TECs, cesium ions can be introduced through various methods, including surface ionization\cite{Rasor_1991_ThermionicEnergyConversion}, impact ionization\cite{Hatsopoulos_1979_ThermionicEnergyConversion, Ito_2012_OpticallyPumpedCesium}, and associative ionization\cite{Pollock_1965_AbsorptionResonanceRadiation, Witting_1965_IonizationProcessLowEnergy}. These methods have been utilized for an extended period, demonstrating stability in both nuclear-heated and combustion-heated systems. Firstly, the surface ionization mechanism, also referred to as contact ionization, involves the ionization of neutral cesium atoms due to the thermionic emission effect and the difference in work function on a metal surface
\cite{Rasor_1991_ThermionicEnergyConversion}. The number of positive ions generated by this process strictly follows the Saha-Langmuir equation\cite{Rasor_1991_ThermionicEnergyConversion}. Secondly, impact ionization technology, also known as electron impact ionization, operates on the principle that high-energy particles collide with neutral cesium atoms, causing them to lose electrons and ionize. This process ultimately results in the formation of a cesium ion and the emission of two electrons\cite{Hatsopoulos_1979_ThermionicEnergyConversion}. Additionally, collisions between electrons and resonantly excited cesium atoms can effectively generate ions for space charge neutralization
\cite{Ito_2012_OpticallyPumpedCesium}. Thirdly, associative ionization refers to the process in which two neutral atoms collide, forming a molecule and releasing an electron, thereby generating ions\cite{Pollock_1965_AbsorptionResonanceRadiation, Nygaard_1973_EffectCesiumPhotoionization, Yamada_1975_DensityProfileCesium}. Notably, associative ionization is recognized as the primary ionization process in low-temperature cesium vapor, as it can directly produce ions without the need for additional energy. This presents a particularly promising outlook for applications in PETE devices\cite{Wang_2024_EffectSpaceCharge}.

In addition to examining the mechanisms for generating positive ions, we are particularly interested in recent advancements in theoretical calculations and simulation studies focused on the space charge effect of positive ions. Within the field of TECs, numerous researchers have conducted in-depth studies on theories related to the calculation of potential distributions influenced by positive ions. In 1960, Auer et al.\cite{Auer_1960_PotentialDistributionsLowPressure} pioneered a method for comprehensively calculating the potential distribution under various operating conditions for a planar diode model of a low-pressure cesium-filled TEC. Subsequently, in 1962, McIntyre et al.\cite{McIntyre_1962_ExtendedSpaceChargeTheory} presented the first integrals of Poisson's equation for six key potential distribution scenarios and obtained numerical solutions using an IBM-704 computer. In 1963, McIntyre et al.\cite{McIntyre_1963_EffectAnodeEmission} further deepened this theory by expanding it to scenarios involving both cathode-emitted ions and electrons, as well as anode-emitted electrons. For the ten most significant potential distribution scenarios, they introduced anode emission contribution terms into the corresponding first-order differential equations for each case. In 1979, Hatsopoulos and Gyftopoulos\cite{Hatsopoulos_1979_ThermionicEnergyConversion} systematically summarized the space charge effect. Building on   the work of Auer et al.\cite{Auer_1960_PotentialDistributionsLowPressure} and McIntyre et al.\cite{McIntyre_1962_ExtendedSpaceChargeTheory}, they\cite{Hatsopoulos_1979_ThermionicEnergyConversion} made several modifications and incorporated them into their book. However, they only covered cases where the collector potential barrier was positive and did not address cases with negative potential barrier. In 2016, Khoshaman et al.\cite{Khoshaman_2016_LowpressurePlasmaenhancedBehavior} developed an algorithm to be capable of self-consistently solving the Vlasov-Poisson equation set to study low-pressure cesium-filled TECs with bidirectional discharges. They found that complex interactions resulting from changes in electron and ion concentrations could lead to phenomena such as plasma oscillations under high ion flux conditions. Additionally, they pointed out that even relatively low ion flux densities could significantly weaken the space charge effect and increase current density. 

In summary, significant progress has been made in the field of TECs with respect to space charge theory, taking into account positive ion input and dual-electrode co-discharges. Since PETE devices exhibit enhanced performance when operated in cesium-filled conditions, they may also be susceptible to intense anode discharges. Therefore, it is crucial to take into account both positive ion neutralization effects and bidirectional discharge effects when analyzing the impact of space charge in PETE devices. In the field of PETE, Ito\cite{Ito_2012_OpticallyPumpedCesium} and Wang et al.\cite{Wang_2024_EffectSpaceCharge} are known pioneers in conducting research on space charge neutralization. Specifically, the PETE model proposed by Wang et al.\cite{Wang_2024_EffectSpaceCharge} in 2024, which incorporates space charge neutralization factors, is groundbreaking but still has certain limitations. These limitations include:

(1) Incomplete consideration of interelectrode motive shape: The positive ion particle number distribution function provided by Wang et al.\cite{Wang_2024_EffectSpaceCharge} only focuses on the case where the anode surface motive is higher than that of the cathode, neglecting the situation where it is lower. This oversight results in the model's output characteristic curves not fully covering all operating voltage ranges.

(2) Absence of electron recycling effect: The model does not account for the electron recycling effect, which has a significant impact on current density and can substantially alter the effective range of the positive ion ratio, thereby affecting the model's accuracy. 

(3) Neglect of reverse discharge impact: The model fails to assess the impact of reverse discharge on the space charge barrier, making it inadequate for evaluating the performance of PETE solar cells containing cesium ions in scenarios with strong anode discharges.

In light of this, this paper innovatively constructs a PIBD (considering positive ion neutralization and bidirectional discharge) model for PETE devices for the first time. This model integrates the combined effects of positive ion injection and dual-electrode discharge on the space charge barrier within the interelectrode space. Importantly, compared to previous models, our PIBD model adopts a more comprehensive space charge theory, allowing the positive ion distribution function to be compatible with both scenarios where the anode motive is higher or lower than the cathode motive. Additionally, it demonstrates applicability in scenarios with strong anode discharges. Based on this, we have conducted precise numerical simulations and detailed parameter analysis to deeply explore the specific impact of strong anode discharges on the space charge neutralization effect. The organizational structure of the subsequent content of this paper is as follows: In Section~\ref{sec:Model}, we establish a theoretical model for space charge neutralization in PETE devices considering bidirectional discharge. In Section~\ref{sec:Result}, we reveal the impact of bidirectional discharge on the neutralization effect through numerical calculations. Finally, we conclude the paper in Section~\ref{sec:Conclusions} by summarizing the key insights obtained from the research.

\section{\label{sec:Model}Theoretical model}

\subsection{The model of photon-enhanced thermionic emission}

\begin{figure*}[htbp]
\centering 
\includegraphics[width=0.8\linewidth]{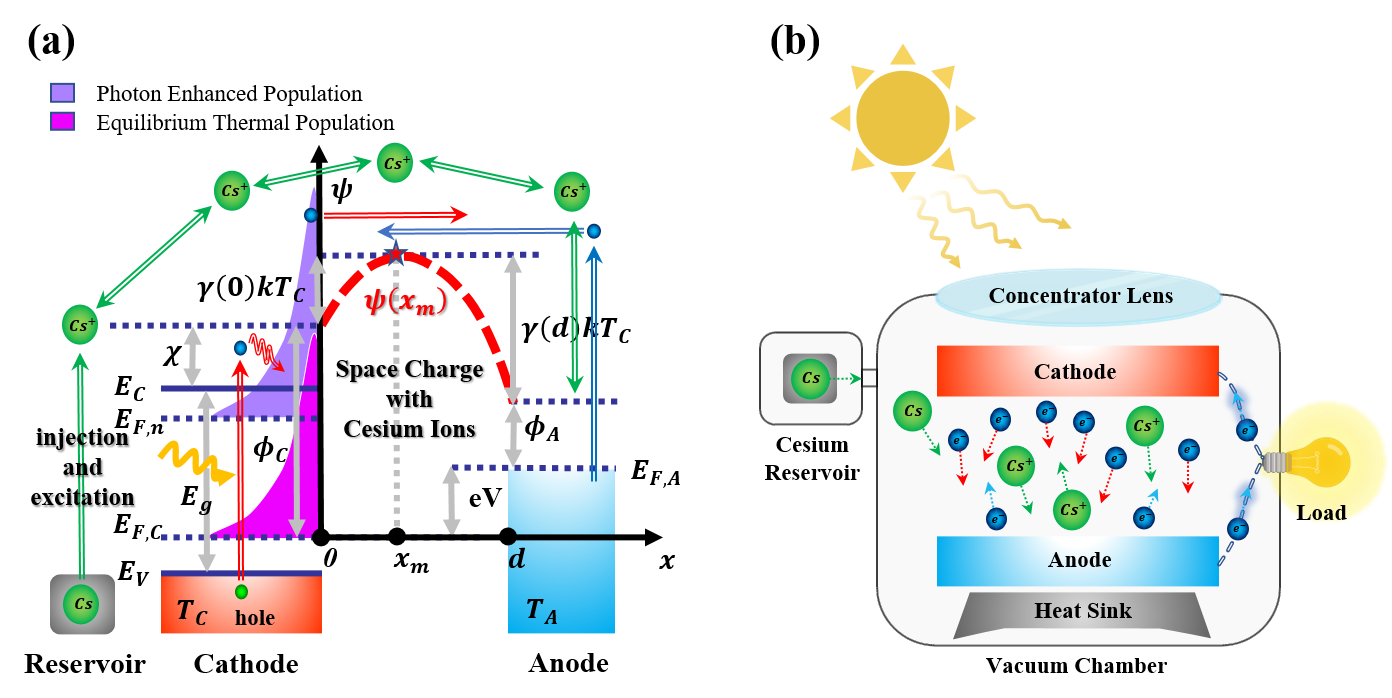} 
\caption{\label{fig:fig_PIBD_schematic}(a) The energy band diagram schematic diagram of a photon-enhanced thermionic emission device considering cesium injection. (b) The implementation of a parallel-plate cesium-filled PETE device that converts solar energy into electrical energy through the photovoltaic effect and thermionic emission.}
\end{figure*}

Fig.~\ref{fig:fig_PIBD_schematic}(a) depicts the energy band diagram schematic diagram of the PETE device. It consists of a p-type semiconductor material acting as the cathode, with an anode metal plate positioned across a vacuum gap.  The red straight arrow emanating from the hole (the green circle) illustrates the excitation of electrons from the valence band to the conduction band upon photon absorption (the yellow curved arrow) . By following a rapid thermalization process, the excited electrons will uniformly disperse throughout the cathode material, achieving an equilibrium distribution in accordance with the cathode's temperature  $T_C$ . Electrons with energies exceeding the electron affinity $\chi$ can be emitted from the cathode surface, thereby generating a thermionic current. As a result, the emitted electrons can leverage the energy of the absorbed photons to surmount the material's bandgap, as well as the thermal energy necessary to overcome the material's electron affinity. When electrons come into the space between the electrodes, they form a space charge barrier. Furthermore, the green straight arrow emerging from the reservoir (positioned next to the cathode within the gray box) signifies both the injection of cesium vapor and its subsequent excitation into cesium ions within the interelectrode space. Once a portion of the vapor is converted into cesium positive ions, these ions, characterized by a distinct velocity distribution, are influenced by an electric field. This force causes them to oscillate reciprocally within the interelectrode gap, as indicated by the remaining green double-headed arrows.

For the two infinite parallel-plate cesium-filled PETE systems illustrated in Fig.~\ref{fig:fig_PIBD_schematic}(b), we assume: 

(1) The cathode C is a semiconductor that can emit photon-excited electrons, while the anode A is a thermal metal also capable of electron emission. The cathode and anode form a parallel-plate structure. Thus, the surface areas available for both photon emission/absorption and electron emission/absorption are equal. Furthermore, the two plates are held at uniform temperatures, denoted as $T_C$ and $T_A$, respectively. Note that the cathode is assumed to be thermally integrated with an ideal solar absorber coating on its surface. This enables the cathode to absorb sub-bandgap photons as heat. Additionally, the cathode is presumed to have a uniform emissivity of unity across the entire spectrum. 

(2) The cesium reservoir R is specifically designed for storing metallic cesium. It is located very close to the horizontal side of the cathode, so we assume that its temperature  $T_R$ is the same as that of the cathode $T_C$. When heated, the metallic cesium in the reservoir R transforms into gaseous cesium, which then evaporates into the adjacent vacuum chamber containing the PETE solar cell. Some of these cesium atoms may become excited and form cesium ions later. 

(3) In the positive \textit{x}-direction, perpendicular to the cathode C, the velocity distribution of electrons emitted from the cathode C and the positive ions released by the cesium reservoir R follows a half-Maxwellian distribution. Similarly, in the negative \textit{x}-direction, the velocity distribution of electrons emitted from the anode A also follows a half-Maxwellian distribution. 

(4) The cesium vapor pressure in the space between the electrodes is sufficiently low, and the number density of cesium atoms is small enough that the effects of particle collisions can be ignored. Additionally, the thermal conduction between the electrodes due to the cesium vapor can be neglected,  and the contribution of electrons generated during cesium atom ionization to the concentration of conduction band electrons in the cathode can also be disregarded.

\subsection{The charge neutrality without illumination}

A p-type semiconductor, specifically boron-doped silicon, serves as the cathode material. The equilibrium Fermi level \(E_{F,C}\) of the semiconductor crystal in the absence of light can be approximated by solving the charge neutrality equation
\begin{equation}
{n_{eq}} + N_A^ -  = {p_{eq}},
\label{eq:neutrality}
\end{equation}
where $n_{eq}$ and $p_{eq}$ represent the equilibrium concentrations of electrons and holes, respectively. These concentrations are given by $n_{eq} = N_C e^{-(E_C - E_{F,C})/kT_C}$ and $p_{eq} = N_V e^{-(E_{F,C} - E_V)/kT_C}$, where $E_C$ and $E_V$ are the energies of the conduction and valence band edges, and $k$ is the Boltzmann constant. The effective densities of states in the conduction and valence bands, $N_C$ and $N_V$, are calculated using $N_C = 2\left(\frac{2\pi m_n^* kT_C}{h^2}\right)^{3/2}$ and $N_V = 2\left(\frac{2\pi m_p^* kT_C}{h^2}\right)^{3/2}$, where \(m_n^* = m_e\) is the effective mass of the electrons, \(m_p^* = 0.57m_e\) is the effective mass of the holes, and \(h\) is Planck's constant. Using the Boltzmann approximation, the concentration of ionized boron acceptors $N_A^- = \frac{N_A }{1 + 4e^{(E_A - E_{F,C})/(kT_C)}}$, where \(E_A = 0.044 \text{ eV}\) is the ionization energy of the boron acceptor in silicon, and \(N_A = 10^{19} \text{ cm}^{-3}\) is the total concentration of boron acceptor impurities.

\subsection{The formulas for particle injection }

\subsubsection{Injection of electrons }

For the PETE device, the expression for the total saturation current density ${J_C}$ emitted out of the cathode is given by\cite{Schwede_2010_PhotonenhancedThermionicEmission}
\begin{equation}
{J_{SC}} = A{T_C}^2{e^{ - \frac{{{\phi _C} - \left( {{E_{F,n}} - {E_{F,C}}} \right)}}{{k{T_C}}}}},
\label{eq:J_SC}
\end{equation}
where the Richardson–Dushman constant \(A = 4\pi em_n^*{k^2}/{h^3}\), ${E_{F,n}}$ is quasi-Fermi level of electron in the cathode, and the work function ${\phi _C} = \chi  + {E_g} - {E_{F,C}}$. The term ${\phi _C} - \left( {{E_{F,n}} - {E_{F,C}}} \right)$ indicates that the effective work function is diminished by the discrepancy between the quasi-Fermi level ${E_{F,n}}$ under photoexcitation and the Fermi level ${E_{F,C}}$ in the absence of photoexcitation.

In non-degenerate semiconductors, the quasi-Fermi level ${E_{F,n}} = {E_{F,C}} + k{T_C}{\rm{ln}}\left( {\frac{n}{{{n_{eq}}}}} \right)$, where $n$ denotes the total electron concentration in the conduction band under photoexcitation conditions. Therefore, it is straightforward to rewrite Eq.~\eqref{eq:neutrality}  as
\begin{equation}
{J_{SC}} = en\sqrt {\frac{{k{T_C}}}{{2\pi {m_n}^*}}} {e^{ - \frac{\chi }{{k{T_C}}}}},
\label{eq:J_SC_2}
\end{equation}
or
\begin{equation}
{J_{SC}} = \frac{n}{n_{eq}}A_RT_C^2\exp\left( - \frac{\phi_C}{kT_C} \right).
\label{eq:J_SC_3}
\end{equation}

For the anode electrode, the saturation current density $J_{SA}$ is calculated by the traditional Richardson equation\cite{Dushman_1923_ElectronEmissionMetals}
\begin{equation}
{J_{SA}} = A{T_A}^2\exp \left( { - \frac{{{\phi _A}}}{{k{T_A}}}} \right),
\label{eq:J_SA}
\end{equation}
where ${\phi _A}$ is the work function of the anode. 

\subsubsection{Injection of positive ions }

Within the scope of this study, our core objective is to revise and improve the space charge theory. Therefore, during numerical simulations, we did not conduct detailed calculations of the cesium ion number density. Instead, we directly adopted the key parameter of cesium ion number density for theoretical calculations and simulations of PETE devices, which can be applied to any type of positive ion injection. Below we will briefly introduce three important ways of producing cesium ions:

(1) Surface ionization. The first ion injection technique, known as surface ionization (or contact ionization), occurs instantaneously when cesium atoms come into contact with a hot cathode. During this process, a cesium ion and an electron are simultaneously generated and released at the cathode surface. To initiate surface ionization, the cesium atoms must absorb energy exceeding their ionization potential. The number of positive ions produced strictly follows the Saha-Langmuir equation and is proportional to the cathode  \(T_R\) and cathode work function \(\phi_R\)\cite{Khoshaman_2016_LowpressurePlasmaenhancedBehavior}. 

(2) Collision ionization. The second ion injection technique is collision ionization (also known as electron-impact ionization), which results in the generation of one cesium ion and two electrons. When the product of the cesium vapor pressure \(P\) and the electrode gap size \(d\) exceeds a certain critical value (i.e., greater than \(3400 \, \mu\text{m} \cdot \text{Pa}\)), the collision effect between the electrons emitted from the cathode and the cesium atoms in the gap becomes significant, triggering collision ionization. Similar to surface ionization, collision ionization also requires the cesium atoms to absorb energy exceeding their ionization potential. Additionally, the number of ions generated by collision ionization is proportional to the cross-sectional area of electron-cesium atom collisions\cite{Wang_2024_EffectSpaceCharge}.

(3) Associative ionization. Although the principles of surface ionization and collision ionization are relatively straightforward, they both require cesium atoms to absorb energy exceeding their ionization potential, which is a rather challenging condition. For example, to ionize cesium atoms, ground-state atoms need to absorb 3.89eV of energy, while the excited-state atoms in the \(6^2P_{1/2}\) and \(6^2P_{3/2}\) level require 2.51eV and 2.44eV, respectively\cite{Wang_2024_EffectSpaceCharge}. Fortunately, excited-state cesium atoms exhibit another ionization mechanism—associative ionization. This mechanism dominates in low-temperature cesium vapor, and when cesium ions are excited to specific excited states, ions can be directly produced without additional energy, making it potentially valuable for applications in PETE devices. The realization of associative ionization involves two key reactions. First, sunlight illuminates the cesium vapor, inducing a photoexcitation process that excites cesium atoms from the ground state \(Cs(6^2 S_{1/2})\) to the first resonant excited energy level \(Cs^*(6^2P_{3/2})\), which can be expressed as
\begin{equation}
Cs(6^2 S_{1/2}) + \frac{hc}{\lambda_0} \rightarrow Cs^*(6^2 P_{3/2}),
\label{eq:Cs reation 1}
\end{equation}
where \(\frac{hc}{\lambda_0}\) ($\approx$1.45 eV) is the photon energy required at the resonant absorption frequency. The \(6^2P_{3/2}\) energy level is chosen because its corresponding photon wavelength is 852.1 nm, which has a high proportion in the solar spectrum and a high absorption efficiency in cesium vapor. Next, excited resonant atoms \(Cs^*(6^2P_{3/2})\) collide and spontaneously form excited cesium molecules \(Cs_2\). These molecules then undergo associative ionization, resulting in the direct production of a molecular ion \(Cs^+_2\) and an electron. This process can be expressed as
\begin{equation}
Cs^*(6^2 P_{3/2}) + Cs^*(6^2P_{3/2}) \rightarrow Cs_2 \rightarrow Cs^+_2 + e^-.
\label{eq:Cs reation 2}
\end{equation}

For specific calculation processes, readers can refer to the appendix of Ref.\cite{Wang_2024_EffectSpaceCharge}.

\subsection{Interelectrode space motive}

\begin{figure*}[htbp]
\centering
\includegraphics[width=0.8\linewidth]{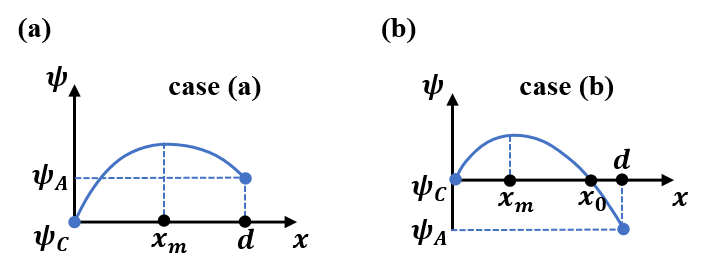} 
\caption{\label{fig:fig_PIBD_motive}(a) The motive diagram when the motive of the anode surface  is greater than 0. (b) The motive diagram when the motive of the anode surface is lower than the cathode's.}
\end{figure*}

When positive ion injection is not considered, the calculation of the electron motive $\psi$ between the electrodes can be performed in a manner similar to that described in the literature\cite{Lin_2024_EffectSpaceCharge}. However, when the positive ion injection is taken into account, we need to consider the contributions of electrons emitted forward from the cathode C, electrons emitted backward from the anode A, and positive ions emitted forward from the cesium reservoir R to the space charge barrier. This requires us to modify the particle distribution functions. The specific approach is as follows: Firstly, the electron distribution functions, $f_{eC}(x, v_e)$ and $f_{eA}(x, v_e)$, are used to describe the electrons emitted from the cathode and anode, respectively, in terms of position $x$ and velocity $v_e$. Similarly, $f_i(x, v_i)$ represents the positive ion distribution function at position $x$ and velocity $ v_i$. Secondly, deriving the charged particle number densities $N_{eC}(x)$ and $N_{eA}(x)$ emitted from the cathode and anode, respectively, as well as the density $N_i(x)$ of positive ions emitted from the cesium reservoir. Then, substittuting these densities into the Poisson equation yields the differential relationship between the motive and the position $x$ in the space between the electrodes. Finally, integrate the Poisson equation to obtain the relationship between $\psi$ and $x$.

The motive $\psi(x)$ is proportional to the electrostatic potential and is determined by solving the Poisson equation in the electrode gap
\begin{equation}
\frac{d^2\psi}{dx^2} = -\frac{e^2}{\epsilon_0}\left[N_{eC}(x) + N_{eA}(x) - N_i(x)\right],
\label{eq:Poisson}
\end{equation}
where $\epsilon_0 = 8.85 \times 10^{-14} \, \text{Fcm}^{-1}$ is the permittivity of vacuum, and the particle number densities at position $x$ can be obtained by integrating the particle distribution function (in Cartesian coordinates) over all velocity components ($v_x$, $v_y$, $v_z$). Specifically

\begin{equation}
N_{eC}(x) = \int_{-\infty}^{\infty} dv_{ex} \int_{-\infty}^{\infty} dv_{ey} \int_{-\infty}^{\infty} dv_{ez} \, f_{eC}(x, v_{ex}, v_{ey}, v_{ez}),
\label{eq:N_eC}
\end{equation}
\begin{equation}
N_{eA}(x) = \int_{-\infty}^{\infty} dv_{ex} \int_{-\infty}^{\infty} dv_{ey} \int_{-\infty}^{\infty} dv_{ez} \, f_{eA}(x, v_{ex}, v_{ey}, v_{ez}),
\label{eq:N_eA}
\end{equation}
and
\begin{equation}
N_i(x) = \int_{-\infty}^{\infty} dv_{ix} \int_{-\infty}^{\infty} dv_{iy} \int_{-\infty}^{\infty} dv_{iz} \, f_i(x, v_{ix}, v_{iy}, v_{iz}).
\label{eq:N_i}
\end{equation}

In practice, each distribution function is a solution to a Vlasov equation. For the one-dimensional geometric model considered, the three Vlasov equations can be written as follows

\begin{equation}
v_{ex} \frac{\partial f_{eC}}{\partial x} - \frac{1}{m_e} \frac{d\psi}{dx} \frac{\partial f_{eC}}{\partial v_{ex}} = 0,
\label{eq:Vlasov_eC}
\end{equation}
\begin{equation}
v_{ex} \frac{\partial f_{eA}}{\partial x} - \frac{1}{m_e} \frac{d\psi}{dx} \frac{\partial f_{eA}}{\partial v_{ex}} = 0,
\label{eq:Vlasov_eA}
\end{equation}
and
\begin{equation}
v_{ix} \frac{\partial f_i}{\partial x} + \frac{1}{m_i} \frac{d\psi}{dx} \frac{\partial f_i}{\partial v_{ix}} = 0,
\label{eq:Vlasov_i}
\end{equation}
where $m_i$ represents the mass of the positive ion.

At the surface of the cathode C, electrons and positive ions are emitted with a half-Maxwellian velocity distribution function characterized by temperature $T_C$. On the other hand, at the surface of the anode A, electrons are also emitted with a half-Maxwellian velocity distribution function characterized by temperature $T_A$. By defining the origin of energy and spatial coordinate at the cathode surface, we obtain

\begin{equation}
{f_{eC}}(0,{v_e}) = 2N_{eC}^+(0){\left( {\frac{{{m_e}}}{{2\pi k{T_C}}}} \right)^{\frac{3}{2}}}\exp \left( { - \frac{{{m_e}v_e^2}}{{2k{T_C}}}} \right)u\left(v_{ex}\right),
\label{eq:f_eC_0}
\end{equation}
\begin{equation}
{f_{eA}}(d,{v_e}) = 2N_{eA}^-(d){\left( {\frac{{{m_e}}}{{2\pi k{T_A}}}} \right)^{\frac{3}{2}}}\exp \left( { - \frac{{{m_e}v_e^2}}{{2k{T_A}}}} \right)u\left(-v_{ex}\right),
\label{eq:f_eA_d}
\end{equation}
and
\begin{equation}
{f_i}\left( {0,{v_i}} \right) = 2N_i^+(0){\left( {\frac{{{m_i}}}{{2\pi k{T_C}}}} \right)^{\frac{3}{2}}}\exp\left( {\frac{{{m_i}v_i^2}}{{2k{T_C}}}} \right)u\left(v_{ix}\right),
\label{eq:f_i_0}
\end{equation}
where $N_{eC}^+(0)$ and $N_i^+(0)$ represent the particle number densities moving in the positive $x$ direction at $x = 0$ (the cathode surface), and $N_{eA}^-(d)$ represents the particle number density moving in the negative $x$ direction at $x = d$ (the anode surface). It is important to note that the particle number density associated with a unidirectional velocity component differs from the actual particle number density, as the latter accounts for particles moving in both the positive and negative $x$ directions.

By taking into account the influence of the space charge barrier, the particle distribution functions at any location can be expressed as

\begin{equation}
\begin{split}
{f_{eC}}(x,{v_e}) &= 2N_{eC}^+(0){\left( {\frac{{{m_e}}}{{2\pi k{T_C}}}} \right)^{\frac{3}{2}}}\exp \left( { - \frac{\psi }{{k{T_C}}} - \frac{{{m_e}v_e^2}}{{2k{T_C}}}} \right) \\ 
&u[{v_{ex}} \pm {(v_{ex})_{min}}],
\label{eq:f_eC_x}
\end{split}
\end{equation}
\begin{equation}
\begin{split}
{f_{eA}}(x,{v_e}) &= 2N_{eA}^+(d){\left( {\frac{{{m_e}}}{{2\pi k{T_A}}}} \right)^{\frac{3}{2}}}\exp \left( { - \frac{\psi-\psi_A}{{k{T_A}}} - \frac{{{m_e}v_e^2}}{{2k{T_A}}}} \right)\\
&u[{-v_{ex}} \mp {(v_{ex})_{min}}],
\label{eq:f_eA_x}
\end{split}
\end{equation}
and
\begin{equation}
\begin{split}
{f_i}\left( {x,{v_i}} \right) &= 2N_i^+(0){\left( {\frac{{{m_i}}}{{2\pi k{T_C}}}} \right)^{\frac{3}{2}}}\exp\left( {\frac{\psi }{{k{T_C}}} - \frac{{{m_i}v_i^2}}{{2k{T_C}}}} \right)\\
&u[{v_{ix}} - (v_{ix})_{min}],
\label{eq:f_i_x}
\end{split}
\end{equation}
where the upper signs in $\pm$ and $\mp$ are used for $x < x_m$ and the lower signs are used for $x \geq x_m$. 

To derive these particle distribution functions in more depth, we first need to clarify the possible ranges of the velocity components $v_{ex}$ and $v_{ix}$ at each location $x$. Notably, these velocity ranges are not arbitrarily set but closely depend on the specific shape of the motive. Therefore, we must determine these velocity ranges one by one in a way that strictly corresponds to the expected shape of the motive.

Specifically, when comprehensively considering the combined effects of forward-emitted electrons, backward-emitted electrons, and forward-emitted positive ions on the space charge barrier, the barrier's morphology may exhibit diversity, necessitating detailed classification and in-depth exploration\cite{McIntyre_1963_EffectAnodeEmission}. However, to address common scenarios in practical applications and simplify the calculation process, we primarily focus on two most typical cases: one where the anode motive is greater than the cathode's ($\psi_A>\psi_C$), and the other where the anode motive is lower than the cathode's ($\psi_A \leqslant \psi_C$). Note that in the subsequent calculations, we set the motive of cathode $\psi_C=0$. 

\begin{figure*}[htbp]
\centering
\includegraphics[width=0.8\linewidth]{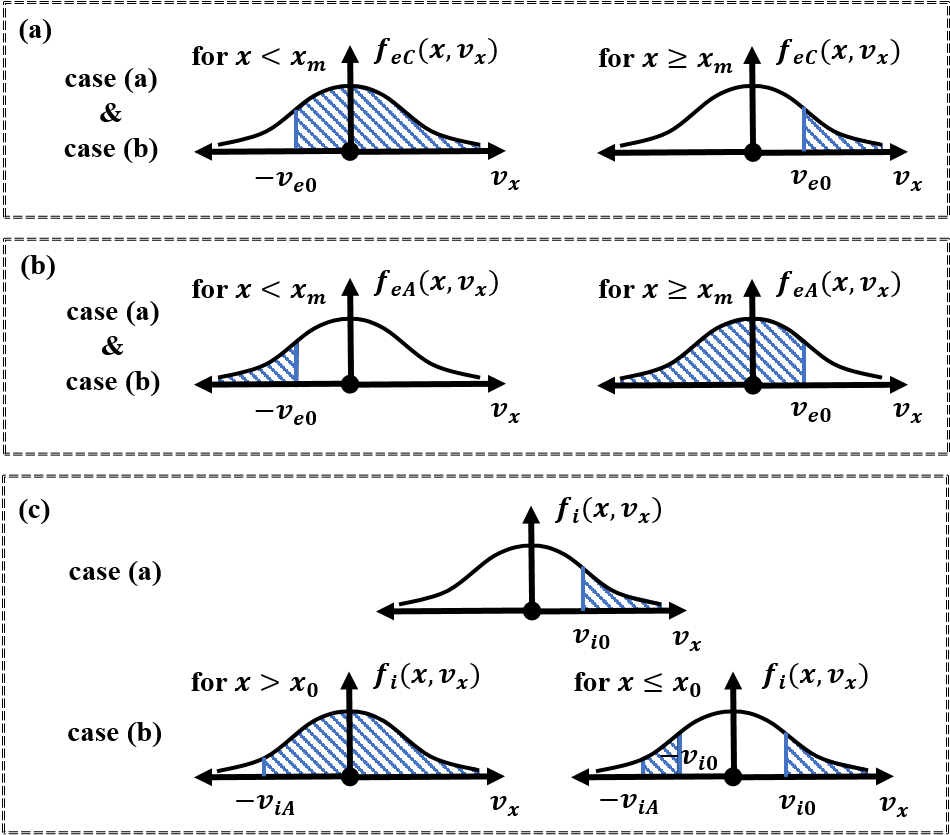} 
\caption{\label{fig:fig_PIBD_velocity}(a) The range of velocity  of the eletrons emitted from the cathode. (b) The range of  velocity of the eletrons emitted from the anode. (c)  The range of velocity  of the ions emitted from the cesium reservoir near the cathode.}
\end{figure*}

\subsubsection{Case (a): the scenario of $\psi_A>\psi_C$}

If the motive is as shown in Figure \ref{fig:fig_PIBD_motive}(a), the motive at points outside the anode is larger than the cathode's ($\psi_A > \psi_C$). Then, according to the discussion in Ref.\cite{Lin_2024_EffectSpaceCharge}, for electrons emitted from the cathode and located at position $x$ and $x < x_m$, the forward minimum velocity can be expressed as
\begin{equation}
v_{e0} = \sqrt{\frac{2(\psi_m - \psi)}{m_i}}.
\label{eq:v_e0}
\end{equation}
When $x \geq x_m$, the minimum reverse  velocity is $-v_{e0}$. Therefore, for electrons emitted from the cathode surface, their velocity range can be expressed as
\begin{equation}
-v_{e0} \leq v_{ex} < \infty \hspace{0.5cm} \text{for} \hspace{0.2cm} x < x_m
\label{eq:eC v_ex <}
\end{equation}
and
\begin{equation}
v_{e0} \leq v_{ex} < \infty \hspace{0.5cm} \text{for} \hspace{0.2cm} x \geq x_m.
\label{eq:eC v_ex >}
\end{equation}

Similarly, for electrons emitted from the anode surface, their velocity range can be expressed as
\begin{equation}
-\infty < v_{ex} \leq -v_{e0} \hspace{0.5cm} \text{for} \hspace{0.2cm} x < x_m
\label{eq:eA v_ex <}
\end{equation}
and
\begin{equation}
-\infty < v_{ex} \leq v_{e0} \hspace{0.5cm} \text{for} \hspace{0.2cm} x \geq x_m.
\label{eq:eA v_ex >}
\end{equation}

Next, the velocity range for positive ions can also be derived using a similar approach to that for electrons. It should be noted that positive ions are decelerated at the region where electrons are accelerated by space charge, and vice versa. Therefore, when the value of $x$ is less than the position of the maximum motive $x_m$, positive ions are accelerated by the electric field and move only in the positive $x$-direction. When $x$ is greater than $x_m$, positive ions are decelerated by the electric field. Because the velocity at  $x$  is determined by the difference in potential energy between the motive at  $x$ and the anode surface, the velocity of the ions at any point is always greater than zero for the scenario of $\psi_A>\psi_C$. Furthermore, since the minimum velocity $v_{i0}$ of positive ions moving in the positive $x$-direction outside the cathode surface is equal to zero, for positive ions at position $x$, the absolute value of their minimum velocity can be expressed as:
\begin{equation}
v_{i0} = \sqrt{\frac{2\psi}{m_i}}.
\label{eq:i v_i0}
\end{equation}

Therefore, when the motive at the anode surface is greater than zero, the velocity range for positive ions can be expressed as
\begin{equation}
v_{i0} \leq v_{ix} < \infty,
\label{eq:i v_ix}
\end{equation}

Next, by integrating the various particle distribution functions from Eqs.~\eqref{eq:f_eC_x}, \eqref{eq:f_eA_x}, and \eqref{eq:f_i_x}, the corresponding particle number density functions can be obtained
\begin{equation}
\begin{split}
N_{eC}(x) &= N_{eC}^+(0)\exp\left(-\gamma\right)\\
&\left[1 \pm \text{erf}\left(\sqrt{\gamma_m - \gamma}\right)\right],
\end{split}
\label{eq:N_eC gamma}
\end{equation}
\begin{equation}
\begin{split}
N_{eA}(x) &= N_{eA}^-(d)\exp\left(\delta\gamma_A - \delta\gamma\right)\\
&\left[1 \mp \text{erf}\left(\sqrt{\delta\gamma_m - \delta\gamma}\right)\right],
\end{split}
\label{eq:N_eA gamma}
\end{equation}
and
\begin{equation}
N_i(x) = N_i^+(0)\exp\left(\gamma\right)\left[1 - \text{erf}\left(\sqrt{\gamma}\right)\right],
\label{eq:N_i gamma psi_A>0}
\end{equation}
where $\text{erf}(x) = \frac{2}{\sqrt{\pi}}\int_0^x \exp(-t^2)dt$, 
\begin{equation}
\gamma = \frac{\psi}{kT_C},
\label{eq:gamma_PIBD}
\end{equation}
and
\begin{equation}
\delta = \frac{T_C}{T_A}.
\label{eq:delta_PIBD}
\end{equation}
In the following discussion, the dimensionless variable $\gamma$ with a single subscript is employed to signify the dimensionless motive at particular locations. For instance, $\gamma_m = \frac{\psi_m}{kT_C}$ and $\gamma_A = \frac{\psi_A}{kT_C}$. For $\gamma$ carries two subscripts, it denotes the dimensionless motive difference between two specific positions, such as $\gamma_{Cm} = \frac{\psi_m - \psi_C}{kT_C}$ and $\gamma_{Am} = \frac{\psi_m - \psi_A}{kT_C}$.

\subsubsection{Case (b): the scenario of $\psi_A \leqslant \psi_C$}

If the motive is as shown in Figure \ref{fig:fig_PIBD_motive}(b),  the motive at the anode's surface is lower than the motive of the cathode's surface ($\psi_A \leqslant \psi_C$). Then, the velocity range for electrons can still be given by the relationship corresponding to case (a) as Eqs.~\eqref{eq:N_eC gamma} and \eqref{eq:N_eA gamma} . However, the velocity range for positive ions differs from the previous scenario. In this case, positive ions with non-zero forward velocities that reach the point $x_0$ at $\psi(x_0)=\psi_C$ will continue moving forward. Under the persistent influence of the reverse electric field, the velocities of  positive ions may decrease to zero and then become negative. 

At any position \(x\), the absolute value of the maximum reverse velocity of the positive ions can be expressed as
\begin{equation}
v_{iA} = \sqrt{\frac{2(\psi - \psi_A)}{m_i}}.
\label{eq:v_iA}
\end{equation}
This value corresponds to the situation where positive ions are just able to reach the anode surface with zero velocity. 

Subsequently, when positive ions are driven backward from the region ($x<x_0$) to the region ($x>x_0$), they will continue to accelerate in the reverse direction due to the electric field, resulting in negative velocities. The maximum reverse velocity, in absolute value, is still given by $v_{iA}$. The minimum reverse velocity in absolute value, can be expressed by Eq.~\eqref{eq:i v_i0} as $v_{i0}$. This corresponds to the scenario where positive ions, initially emitted from the cathode with zero velocity, reach $x_0$ with also zero velocity, and then are accelerated reversely.

In summary, when the motive at the anode surface is lower than the cathode's, the velocity distribution of positive ions can be written as follows

\begin{equation}
\begin{cases}
-v_{iA}\leqslant v_{ix} < \infty  \hspace{0.5cm}
&\text{for} \hspace{0.2cm} x > x_0 \\
(-v_{iA} \leqslant v_{ix} < -v_{i0}) \cup (v_{i0} \leqslant v_{ix} < \infty) \hspace{0.5cm}
&\text{for} \hspace{0.2cm} x \leqslant x_0
\end{cases}.
\label{eq:v_ix <>}
\end{equation}

Next, by integrating the respective particle distribution functions, we can derive the corresponding particle number density functions for positive ions as
\begin{equation}
\begin{split}
N_i(x) &= N_i^+(0) \exp(\gamma) 
\left[ \text{erf}\left( \sqrt{\gamma - \gamma_A} \right) \right. \\
&\quad + \left.
\begin{cases}
  1 & \text{for } \hspace{0.2cm} \gamma < 0 \\
  1 - 2 \, \text{erf}(\sqrt{\gamma}) & \text{for} \hspace{0.2cm} \gamma \geq 0
\end{cases}
\right]
\end{split},
\label{eq:N_i gamma psi_A<0}
\end{equation}
which differs from Eq.~\eqref{eq:N_i gamma psi_A>0}.

\subsubsection{Dimensionless Poisson Equation}

By substituting various particle number density functions into the Poisson equation, followed by dimensionless the horizontal and vertical coordinates, and then performing a series of simplifications, we obtain the dimensionless Poisson equation
\begin{equation}
\frac{d^2\gamma}{d\xi^2} = -\frac{1}{2}\left[ n_{eC}(\xi) + \beta n_{eA}(\xi) - \alpha n_i(\xi) \right],
\label{eq:possion dimensionless}
\end{equation}
where $\xi = {x}/{\sqrt{\frac{\varepsilon_0 kT_C}{2e^2N_{eC}^+(0)}}}$, the proportional coefficients are
\begin{equation}
\begin{cases}
\alpha = \frac{N_i^+(0)}{N_{eC}^+(0)} \\
\beta = \frac{N_{eA}^-(\xi_A)}{N_{eC}^+(0)}
\end{cases},
\label{eq:proportional coefficients}
\end{equation}
and the reduced particle number densities are
\begin{equation}
\begin{cases}
n_{eC}(\xi) = \frac{N_{eC}(\xi)}{N_{eC}^+(0)} \\
n_{eA}(\xi) = \frac{N_{eA}(\xi)}{N_{eA}^-(\xi_A)} \\
n_i(\xi) = \frac{N_i(\xi)}{N_i^+(0)}
\end{cases}.
\label{eq:reduced}
\end{equation}

Here, $\xi_A$ represents the dimensionless horizontal coordinate at the anode surface, corresponding to the position $x = d$. It can be observed that when the coefficient $\alpha = 0$, the equation reduces to the form presented in Ref.\cite{Lin_2024_EffectSpaceCharge}; and when both $\alpha = 0$ and $\beta = 0$, the equation further reduces to the form presented in Ref.\cite{Su_2014_SpaceChargeEffects}.

\subsubsection{Proportional coefficients}

To solve the aforementioned dimensionless differential equation, the values of two proportional coefficients are required. Firstly, by integrating the particle distribution functions with respect to velocity in one direction, the cathode and anode saturation current densities can be obtained respectively
\begin{equation}
J_{SC} = eN_{eC}^+(0)\sqrt{\frac{2kT_C}{\pi m_e}} = J_C\exp(\gamma_{Cm})
\label{eq:J_SC N}
\end{equation}
and
\begin{equation}
J_{SA} = eN_{eA}^-(d)\sqrt{\frac{2kT_A}{\pi m_e}} = J_A\exp(\gamma_{Am}).
\label{eq:J_SA N}
\end{equation}

Thus, for the positive ion richness ratio $\alpha$, we have
\begin{equation}
\alpha = \frac{N_i^+(0)}{N_{eC}^+(0)} = N_i^+(0)\frac{e}{J_{CS}}\sqrt{\frac{2kT_C}{\pi m_e}}.
\label{eq:alpha}
\end{equation}

In particular, during the actual calculations in this paper, we pre-determine the positive ion richness ratio $\alpha$ and then calculate $N_i^+(0)$ based on it, aiming to explore the impact of the positive ion richness ratio on space charge neutralization effects. Furthermore, for $\beta$, we have
\begin{equation}
\beta = \frac{N_{eA}^-(d)}{N_{eC}^+(0)} = \frac{J_{AS}\sqrt{T_C}}{\sqrt{T_A}J_{CS}}.
\label{eq:PIBD beta}
\end{equation}

Notably, when calculating the proportional coefficient $\beta$ in this paper, we first estimate the four parameters of the PETE device, use them to compute $\beta$, and then substitute it into the Poisson equation for a "self-consistent iterative" solution.

By combing Eq.~\eqref{eq:J_SC},  ~\eqref{eq:J_SA}, ~\eqref{eq:J_SC N} and ~\eqref{eq:J_SA N}, we can obtain the values of coefficients $\alpha$ and $\beta$, and then determine the  the dimensionless motive from Eq.~\eqref{eq:possion dimensionless}. Then, the net current density $J$ of the PETE device can be obtained as
\begin{equation}
\begin{split}
J &= {J_C} - {J_A} \\
&= AT_C^2{e^{ - \frac{{{\phi _C} - \left( {{E_{F,n}} - {E_{F,C}}} \right) + \gamma_{Cm}k{T_C}}}{{k{T_C}}}}} - AT_A^2{e^{ - \frac{{{\phi _A} + \gamma_{Am}k{T_C}}}{{k{T_A}}}}}.
\end{split}
\label{eq:J=J_C-J_A}
\end{equation}

\subsection{The electron continuity equation in the conduction band}

Equation~\eqref{eq:J=J_C-J_A} reveals that the net current density hinges on the quasi-Fermi level ${E_{F,n}}$ of electrons within the cathode. Prior to computing ${E_{F,n}}$, it is essential to ascertain the electron concentration $n$ by addressing the electron continuity equation within the conduction band, formulated as
\begin{equation}
{\Gamma _{Sun}} - {\Gamma _R} = \frac{{{J_C} - {J_A}}}{eL},
\label{eq:electron continuity}
\end{equation}
where the electron generation rate ${\Gamma _{Sun}}$ is defined as $\frac{{\Phi _{Sun}}\left( {E > {E_g}} \right)}{L}$\cite{Schwede_2010_PhotonenhancedThermionicEmission}, ${\Phi _{Sun}}\left( {E > {E_g}} \right)$ signifies the photon flux density exceeding the bandgap energy in the AM1.5 Direct (+circumsolar) solar spectrum with concentration, and $L$ denotes the film thickness. The photon-enhanced radiative recombination rate ${\Gamma _R}$ is given by $\frac{1}{L}\left( \frac{np}{{{n_{eq}}{p_{eq}}}} - 1 \right)\frac{{2\pi }}{{{h^3}{c^2}}}\int_{{E_g}}^\infty  \frac{{{{(h\nu )}^2}d\left( {h\nu } \right)}}{{{e^{h\nu /\left( {k{T_C}} \right)}} - 1}}$\cite{Schwede_2010_PhotonenhancedThermionicEmission}, involving the speed of light $c$, the hole concentration $p$ in the valence band influenced by photoexcitation, and the photon energy $h\nu$. Within this analysis, radiative recombination stands as the sole recombination mechanism considered. Analogous to Ref.\cite{Segev_2015_NegativeSpaceCharge, Lin_2024_EffectSpaceCharge}, Eq.~\eqref{eq:electron continuity} presumes that both reverse-emitted electrons and electrons reflected by the energy barrier in the inter-electrode space contribute to the electron population (note that the number of electrons generated when cesium atoms generate cesium ions is smaller than that of the above parts, so it is not considered), thereby illustrating the electron recycling effect. The detailed derivation of $n$ from Eq.~\eqref{eq:electron continuity} is provided in the supplementary materials of Ref.\cite{Segev_2015_NegativeSpaceCharge}.

\subsection{The energy balance at the cathode}

Additionally, we examine the energy balance pertinent to the cathode. We disregard the radiative heat transfer between the cathode and anode by presuming the anode to be fully reflective. By integrating an infrared (IR) coupling element within the cathode to capture sub-bandgap photons as heat, the energy balance equation for the cathode is formulated as
\begin{equation}
\begin{split}
P_{\text{sun}} &= P_{\text{IR}} + P_0 + P_{\text{rad}} \\
&\quad + J_C\left[ \psi_m + 2kT_C \right] - J_A\left[ \psi_m + 2kT_A \right],
\end{split}
\label{eq:energy balance}
\end{equation}
where $P_{\text{sun}}$ signifies the input energy density from the concentrated AM1.5 Direct (+circumsolar) spectrum. $P_0 = \frac{2\pi}{h^3c^2}\int_{E_g}^\infty \frac{(hv)^3 d(hv)}{e^{hv/(kT_C)} - 1}$ represents the thermal energy loss stemming from equilibrium radiative recombination within the cathode. $P_{\text{rad}} = P_0\left[ \frac{np}{n_{\text{eq}}p_{\text{eq}}} - 1 \right]$ accounts for the thermal energy loss due to non-equilibrium radiative recombination. Furthermore, $P_{\text{IR}} = \frac{2\pi}{h^3c^2}\int_0^{E_g} \frac{(hv)^3 d(hv)}{e^{hv/(kT_C)} - 1}$ denotes the energy loss resulting from radiation emitted by the IR coupling element. 

It is worth noting that the IR coupling element absorbs all sub-bandgap radiation and remains transparent to supra-bandgap photons. The combined $P_{\text{IR}}$ and $P_0$ encapsulate the full-spectrum blackbody emission of the cathode. The last two terms in Eq.~\eqref{eq:energy balance} indicate the thermal energy transported by electrons emitted from both the cathode and anode. Each electron emitted from the cathode to the anode conveys an energy of $\psi_m + 2kT_C$, whereas each electron transitioning from the anode to the cathode surrenders an energy of $\psi_m + 2kT_A$ upon arrival\cite{Hatsopoulos_1973_ThermionicEnergyConversion, ODwyer_2007_SolidstateRefrigerationPowergeneration}.

\subsection{The power output and efficiency }

Finally, the voltage in the space charge limited regime can be written as 
\begin{equation}
V = \left\{ {\left[ {{\phi _C} + \gamma_{Cm} k{T_C}} \right] - \left[ {{\phi _A} + \gamma_{Am} k{T_C}} \right]} \right\}/e.
\label{eq:V}
\end{equation}
The power output and efficiency of the PETE device can be expressed as
\begin{equation}
P = JV,
\label{eq:P}
\end{equation}
and
\begin{equation}
\eta = JV/{P_{sun}}.
\label{eq:eta}
\end{equation}

\section{\label{sec:Result}Results and discussion}

\begin{figure*}[htbp]
\centering
\includegraphics[width=0.9\linewidth]{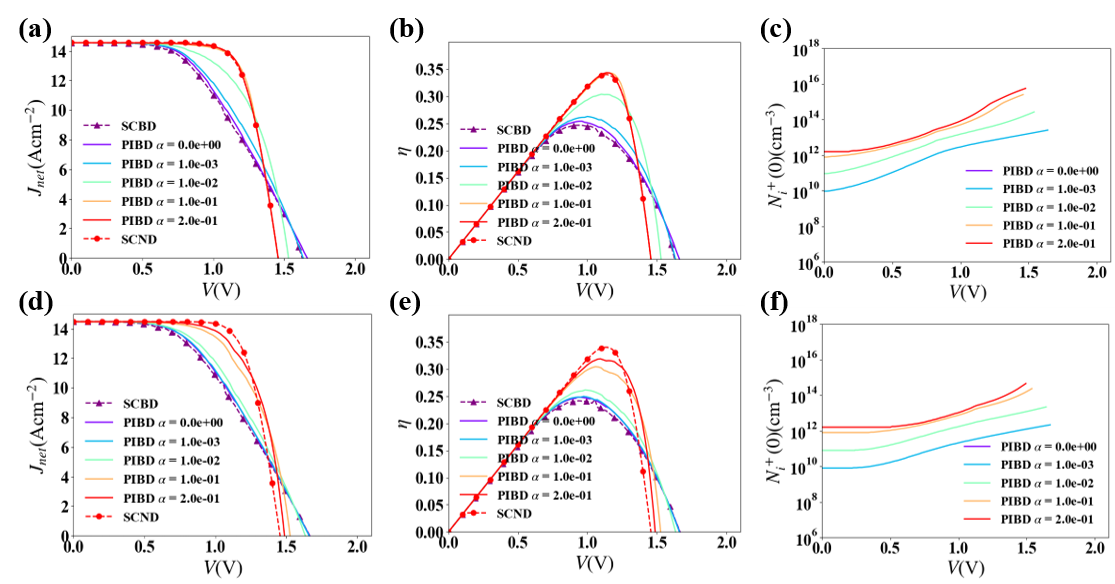} 
\caption{\label{fig:fig_PIBD_output}The figure shows the influence of the positive ion richness ratio $\alpha$ on the output characteristics of the PETE device. The electron affinity $\chi$ is assigned a value of 0.6eV for the first row of the figure and 0.9eV for the second row. The first column of the figure shows the $J_{net}-V$ curves, the second column displays the $\eta-V$ curves, and the third column presents the $N_i^+(0)-V$ curves.}
\end{figure*}

In the following simulation, unless otherwise explicitly stated, we typically use  the  parameters: the solar concentration ratio $C$ is set to 500, the anode temperature $T_A$ is set to 600K, the interelectrode gap $d$ is fixed at 5$\rm{\mu m}$, the bandgap $E_g$ is selected as 1.4V, the electron affinity $\chi$ is determined to be 0.6eV, and the anode work function $\phi_A$ is set to 0.9eV.  The cathode temperature $T_C$ is determined by Eq.~\eqref{eq:energy balance}, As for the positive ion richness ratio $\alpha$, we establish its value range between $10^{-3}$ and $6 \times 10^{-1}$ to facilitate detailed analysis and discussion.

\subsection{The influence of the positive ion richness ratio}

Now, let's discuss the impact of the positive ion richness ratio $\alpha$ on the output characteristics of PETE devices. As shown in Fig.~\ref{fig:fig_PIBD_output}, the purple dashed-dotted line depicts the SCBD (space charge effect due to the bidirectional discharge, but without considering Cesium ion neutralizatio) model, the red dashed-dotted line represents the SCND (without considering the space charge effect due to the discharge) model, and the solid lines in various rainbow colors collectively illustrate different scenarios of the PIBD (space charge effect simultaneously considering Cesium ion neutralization and bidirectional ischarge) model. From a theoretical perspective, when the positive ion richness ratio $\alpha$ equals 0, the PIBD model fully degrades to the SCBD model, as the number of positive ions is zero, and the ability to neutralize the space charge cannot be achieved. Conversely, as $\alpha$ approaches 1 infinitely, the PIBD model infinitely approximates the SCND model. At this point, the number of positive ions is sufficiently large compared to electrons,  which is sufficient to completely counteract the space charge potential barrier generated by electrons. This effect may  completely eliminate the inhibitory influence of space charge on electron motion. 

It's worth mentioning that compared to the model proposed by Wang et al.\cite{Wang_2024_EffectSpaceCharge}, our model provides a more comprehensive description of the space charge potential barrier function. Specifically, our model not only considers the case where the anode surface motive $\psi_A$ is greater than the cathode's but also accounts for scenarios where $\psi_A$ is lower than cathode's. This comprehensive consideration is indispensable when conducting numerical simulations to comprehensively cover all operating voltage $V$ ranges. Furthermore, this improvement introduces some special curve variations, as shown in Fig.~\ref{fig:fig_PIBD_output}, where subtle fluctuations (induced by the modified potential barrier function) appear in the curves of net current density $J_{net}$ and particle number density $N_{i}^+(0)$. The reason is that as we gradually increase the operating voltage $V$ to calculate the output characteristics, $\psi_A$ also increases accordingly. During this process, once $\psi_A$ changes from negative to positive, one part of the Poisson equation undergoes a sudden change (from Eq.~\eqref{eq:N_i gamma psi_A<0} to Eq.~\eqref{eq:N_i gamma psi_A>0} ), causing the solution of the differential equation to be discontinuous at that point, which in turn results in abrupt slope changes in various output characteristic curves at the corresponding locations. For instance, as illustrated in Figs.~\ref{fig:fig_PIBD_output} (d) and (e), the $J-V$ curve or the $\eta-V$ curve exhibits a sudden upward protrusion at a certain point.

\begin{figure*}[htbp]
\centering
\includegraphics[width=0.9\linewidth]{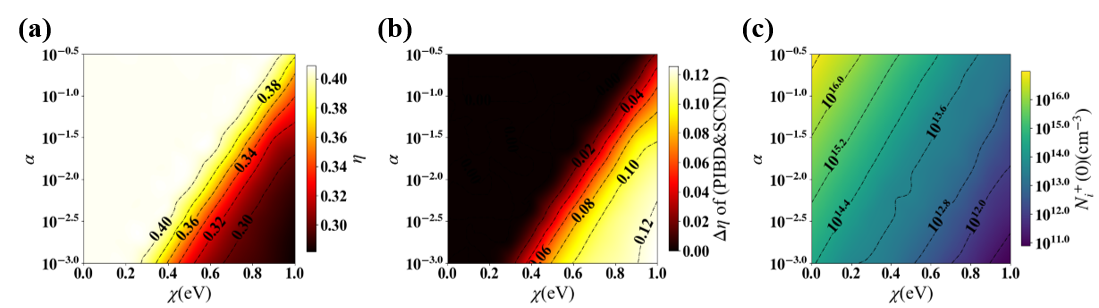} 
\caption{\label{fig:fig_PIBD_affinity}The contour plots of (a) the efficiency $\eta$ of the PIBD model, (b) deviation $\Delta\eta$ of the efficiency of the PIBD model with that of the SCND model, and (c)  ion number density $N_i^+(0)$ varying with the electron affinity $\chi$ and the ion-richness ratio $\alpha$, where the voltage $V$ has been optimized.}
\end{figure*}

\subsection{The influence of the electron affinity}

Now, let's discuss the impact of electron affinity $\chi$ on the effective range of the positive ion abundance ratio $\alpha$. As shown in Fig.~\ref{fig:fig_PIBD_output}(a), under the conditions of a solar concentration ratio $C = 500$ and an electron affinity $\chi = 0.6$eV, as $\alpha$ increases from 0 to $10^{-3}$, the curve of the PIBD model gradually diverges from the SCBD model, indicating that the space charge effect begins to be significantly counteracted by positive ions. When $\alpha$ further increases to $10^{-1}$, the curve of the PIBD model gradually approaches the SCND model, suggesting that the space charge effect is nearly completely neutralized by positive ions. In this context, if we define the "effective range" of the positive ion richness ratio $\alpha$ as the parameter range where the PIBD model begins to significantly diverge from both the SCBD and SCND models, we can conclude that the effective range of $\alpha$ at this point is roughly from $1 \times 10^{-3}$ to $1 \times 10^{-1}$.

Next, if we consider a larger electron affinity $\chi$, we find that the effective range of $\alpha$ expands accordingly. For example, as shown in Fig.~\ref{fig:fig_PIBD_output}(b), when $\chi$ increases to $0.9$eV, the effective range of $\alpha$ expands to the region between $1 \times 10^{-2}$ and $2 \times 10^{-1}$. The underlying reason for this phenomenon is that our model incorporates the effect of electron recycling. Specifically, when the electron affinity $\chi$ is small, the cathode current density $J_C$ is relatively large, resulting in a higher electron number density $N_{eC}^+(0)$ at the cathode surface (contributed by the cathode). At the same $\alpha$ value, the positive ion number density $N_{i}^+(0)$ at the cathode surface is also correspondingly higher, providing a stronger neutralization capability for space charge and making it easier to approach the effects of the SCND model. Conversely, when the electron affinity $\chi$ is large, the cathode current density $J_C$ decreases, causing $N_{eC}^+(0)$ to decrease as well. This in turn leads to a reduction in $N_{i}^+(0)$, making it more difficult to achieve the effects of the SCND model.

As clearly observed in Fig.~\ref{fig:fig_PIBD_affinity}(a), the efficiency $\eta$ of PETE devices exhibits a significant upward trend as the positive ion abundance ratio $\alpha$ gradually increases. This phenomenon is attributed to the fact that a higher $\alpha$ value implies an increased positive ion density, which can more effectively neutralize the space charge barrier formed by electrons between the electrodes. Conversely, when the electron affinity $\chi$ increases, the efficiency of PETE devices tends to decrease. This is mainly because, after considering the electron recycling effect, an increase in $\chi$ hinders the escape of electrons from the cathode surface, resulting in a reduced current density emitted by the cathode.

Furthermore, by closely examining Fig.~\ref{fig:fig_PIBD_affinity}(b), we can further reveal the intrinsic relationship between the electron affinity $\chi$ and the positive ion abundance ratio $\alpha$ required to achieve complete "space charge neutrality." Specifically, when the electron affinity $\chi$ is at a lower level, the positive ion richness ratio $\alpha$ needed for the PIBD model to align with the SCND model in terms of the output characteristics (i.e., the efficiency difference $\Delta\eta$ approaching 0) is also relatively low. Conversely, if the electron affinity $\chi$ is higher, the required $\alpha$ value will increase accordingly. The root of this correlation lies in the fact that a decrease in electron affinity $\chi$ promotes an increase in the cathode saturation current density $J_{SC}$. Under constant $\alpha$ conditions, this change directly leads to an increase in the positive ion density $N_i^{+}(0)$ at the anode surface [shown in Fig.~\ref{fig:fig_PIBD_affinity}(c)], and subsequently enhances the positive ion density $N_i(x)$ at other locations in space, thereby more effectively eliminating the space charge barrier between the electrodes.

\begin{figure*}[htbp]
\centering
\includegraphics[width=0.9\linewidth]{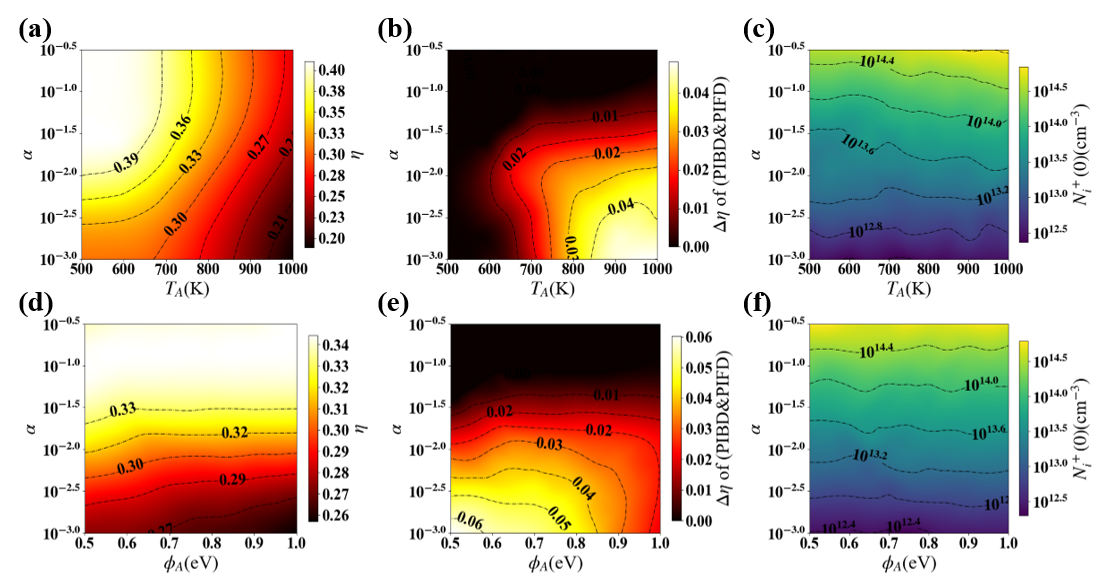} 
\caption{\label{fig:fig_PIBD_anode}The contour plots of (a) the efficiency $\eta$ of the PIBD model, (b) deviation $\Delta\eta$ of the efficiency of the PIBD model with that of the PIFD model, and (c)  ion number density $N_i^+(0)$ varying with the anode temperature $T_A$ and the ion-richness ratio $\alpha$, where the voltage $V$ has been optimized. In subgraphs (a-c), $\phi_A$ is set to 0.9eV, and in subgraphs (e-f), $T_A$ is set to 800K.}
\end{figure*}

\subsection{The influence of the bidirectional discharge}

Next, we conducted an in-depth study on the specific impact of the anode temperature on the PETE performance. As shown in Fig.~\ref{fig:fig_PIBD_anode}(a), it is clearly observable that as the anode temperature $T_A$ increases, the efficiency $\eta$ of the PETE devices exhibits a downward trend. Meanwhile, Fig.~\ref{fig:fig_PIBD_anode}(b) further reveals that an increase in the anode temperature also results in a significant enlargement of the efficiency deviation $\Delta\eta$ between the PIBD model and the PIFD model. The reason for this phenomenon lies in the fact that the PIBD model incorporates the influence of anode discharge on the space charge barrier. When the anode temperature rises, the anode discharge effect intensifies. It will increase the number of electrons emitted by the anode, which subsequently exacerbates the formation of the space charge barrier between the electrodes. It will aso hinder the smooth arrival of electrons emitted by the cathode to the anode. Additionally, a higher anode temperature implies that more positive ions are required to achieve efficiency saturation, i.e., the required positive ion richness ratio $\alpha$ also increases. To neutralize the additional space charge barrier formed by electrons emitted from the anode, the system has to introduce more positive ions to maintain balance. Specifically, Fig.~\ref{fig:fig_PIBD_anode}(c) provides a detailed presentation of the positive ion number density $N_i^+(0)$ under different parameter conditions.

We also delved into the specific impact of the node work function on the PETE performance. As shown in Fig.~\ref{fig:fig_PIBD_anode}(e), it is evident that a decrease in the anode work function $\phi_A$ leads to an improvement in the efficiency $\eta$ of PETE devices. This is because a smaller anode work function results in a larger flat-band voltage $V_{flat}$ ($=\phi_C-\phi_A$), causing the maximum efficiency point $\eta_{max}$ of PETE devices to occur in a region of higher operating voltage $V$. However, as shown in Fig.~\ref{fig:fig_PIBD_anode}(f), a reduction in the anode work function also leads to an increase in the efficiency deviation $\Delta\eta$ between the PIBD model and the PIFD model. The underlying reason is that when the PIBD model considers the contribution of the anode discharge to the space charge barrier, a decrease in anode work function enhances the anode discharge capability. It results in an increased number of electrons emitted by the anode, which subsequently exacerbates the formation of the space charge barrier between the electrodes, and hinders the smooth arrival of electrons emitted by the cathode to the anode. Furthermore, due to the need to balance the efficiency gain from increasing the flat-band voltage and the efficiency loss caused by anode discharge, the relationship between the anode work function and the positive ion richness ratio $\alpha$ required to achieve efficiency saturation becomes relatively complex, but generally exhibits a positive correlation trend. Specifically, Fig.~\ref{fig:fig_PIBD_anode}(c) provides a detailed presentation of the positive ion number density $N_i^+(0)$ under different parameter conditions.

\section{\label{sec:Conclusions}Conclusions}
In this study, we explore the space effects in PETE devices by examining the contributions of positive ion injection and bidirectional discharge to the space charge barrier in the interelectrode region. Through numerical simulations, we carefully assessed the influence of key parameters. The findings indicate that  smaller values of $\alpha$ lead to more significant space charge effects. Accounting for the electron recycling effect, we observed that a lower electron affinity $\chi$ reduces the magnitude of $\alpha$ required for complete space charge neutralization.  Furthermore, it is observed that a higher anode temperature or a lower anode work function amplifies the effect of reverse discharge on net current density. In contrast, a higher anode temperature or a greater anode work function increases the proportion of positive ions needed for complete space charge neutralization. In summary, through theoretical derivation and numerical simulations, our study elucidates the mechanisms and characteristics of the space charge neutralization effect in PETE devices, establishing a foundational theory that can enhance their design. Furthermore, the corresponding theory and algorithm hold potential to inspire research endeavors in various other fields.


\begin{acknowledgments}
This work has been supported by the National Natural Science Foundation of China (12364008), Natural Science Foundation of Fujian Province (2023J01006), and Fundamental Research Fund for the Central Universities (20720240145, 20720230012).
\end{acknowledgments}


\appendix

\section{\label{sec:numerical}Design of numerical procedure}

Before introducing the numerical procedures, we need to define the different operating points and regions of a PETE device:
(1) Firstly, we define the operating voltage range that causes the maximum motive $\psi_m$ to appear at the cathode surface as the "saturation region," and the operating voltage that just achieves this state is defined as the saturation voltage $V_{Sat}$.
(2) Secondly, we define the operating voltage range that causes the maximum motive $\psi_m$ to appear at the anode surface as the "retrading region," and the operating voltage that just achieves this state is defined as the critical voltage $V_{Cri}$ (note that to obtain $V_{Cri}$, the subsequent "space-charge limited region" calculation needs to be performed first).
(3) Finally, we define the operating voltage range that causes the maximum motive $\psi_m$ to appear between the cathode surface and the anode surface as the "space-charge limited region," where the operating voltage $V$ lies between $V_{Sat}$ and $V_{Cri}$.

In this paper, it should be noted that to make the electron concentration $n$ in the cathode conduction band vary with the operating voltage $V$, the particle rate balance equation \eqref{eq:electron continuity} of PETE must be coupled with the negative space charge model proposed in this paper. Additionally, since we consider energy balance, equation \eqref{eq:energy balance} should also be coupled with the aforementioned set of equations.

It is worth noting that, for convenience in representing variables, we have defined several relative values—$\psi_{ij}$, $\gamma_{ij}$, and $\xi_{ij}$—for the motive $\psi$, the dimensionless barrier $\gamma$, and the dimensionless position $\xi$, respectively. Here, $i$ or $j$ can represent any one of C (cathode), A (anode), and m (maximum motive). These relative values can be used to quantify the differences in corresponding quantities between two different positions. 

Specifically, after taking into account both positive ion injection and bidirectional space charge effects, we define the zero points of $\gamma$ and $\xi$ at the vacuum level of the cathode surface, rather than at the maximum motive potential $\psi_m$ as reported in Ref.\cite{Lin_2024_EffectSpaceCharge}. This adjustment is made to better align the dimensionless variables with the space charge theory presented in this paper. Furthermore, the object participating in the "self-consistent iterative" algorithm in the space-charge limited region has also changed. For instance, the object that needs to be checked for convergence shifts from $\psi_m$ to $\xi_0$.

Next, we will elaborate on the calculation methods for each region in detail.

\subsection{The algorithm of the saturation region}

\begin{figure*}[htbp]
\centering 
\includegraphics[width=0.8\linewidth]{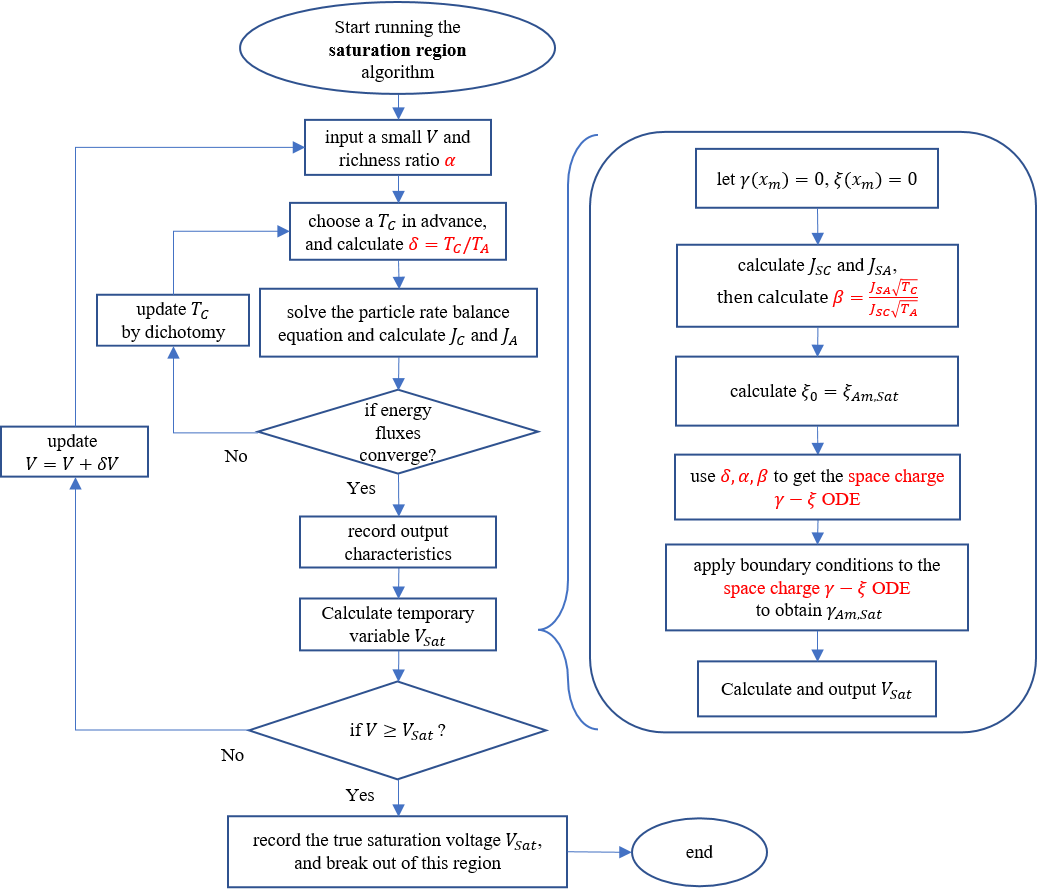} 
\caption{\label{fig:fig_PIBD_algor1}The flow chart of saturation region and saturation point algorithm.}
\end{figure*}

\begin{figure*}[htbp]
\centering 
\includegraphics[width=0.8\linewidth]{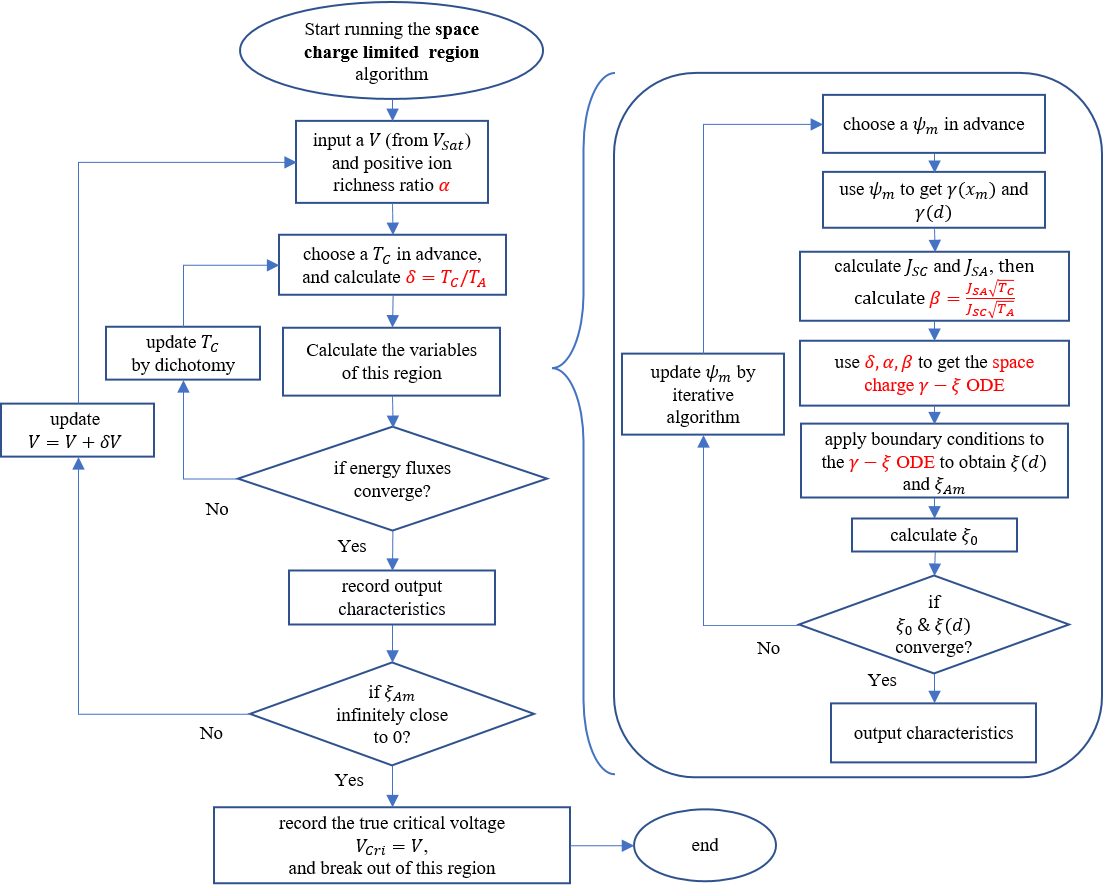} 
\caption{\label{fig:fig_PIBD_algor2}The flow chart of space-charge limiter region and critical point algorithm.}
\end{figure*}

In the saturation region, the maximum motive $\psi_m$ appears at the cathode surface. At this point, we have ${\psi _{Cm}} = 0$.  Therefore,  the cathode current density ${J_C}$ equals the cathode saturation current density, given by ${J_C} = \frac{n}{{{n_{eq}}}}{A_R}{T_C}^2\exp\left( { - \frac{{{\phi _C}}}{{k{T_C}}}} \right)$, while the anode current density can be expressed as ${J_A} = {A_R}{T_A}^2\exp\left( { - \frac{{{\phi _C} - eV}}{{{k_B}{T_A}}}} \right)$.

Due to the introduction of the electron recycling effect, the calculated $V_{Sat}$ may vary for different $V$. Therefore, in the saturation region, we will presuppose a small operating voltage $V$ as the lower bound. Subsequently, we will incrementally increase $V$ and calculate the corresponding output characteristics, while simultaneously monitoring in real-time whether the saturation point has been reached. The specific algorithm flow is as follows:

1. Select an operating voltage $V$ (starting from a small value and incrementally increasing it) and a positive ion richness ratio $\alpha$, and begin the calculation.

2. Calculate the cathode and anode current densities ${J_C} = \frac{n}{{{n_{eq}}}}{A_R}{T_C}^2\exp\left( { - \frac{{{\phi _C}}}{{k{T_C}}}} \right)$ and ${J_A} = {A_R}{T_A}^2\exp\left( { - \frac{{{\phi _C} - eV}}{{{k_B}{T_A}}}} \right)$ based on the solution of the particle rate balance equation \eqref{eq:electron continuity} for the current operating voltage $V$.

3. Repeat step 2, using the energy balance equation \eqref{eq:energy balance} to determine $T_C$ and equation \eqref{eq:delta_PIBD} to determine $\delta$.

4. Set the value of the dimensionless barrier $\gamma(x_m)$ at the maximum barrier location to 0, and the dimensionless distance $\xi(x_m)$ also to 0.

5. Calculate the cathode and anode saturation current densities $J_{SC}$ and $J_{SA}$, and then determine $\beta$ using equation \eqref{eq:PIBD beta}.

6. Calculate the dimensionless distance $\xi_0$ between the two electrodes (at this point, its value is equal to the dimensionless distance $\xi_{Am,Sat}$ between the anode and the maximum barrier at the saturation point):
\begin{equation}
{\xi_0} = {\left( {\frac{{2\pi {m_e}{e^2}}}{{\varepsilon _{_0}^2{k^3}}}} \right)^{0.25}}\frac{{{J_C}^{0.5}d}}{{T_C^{0.75}}};
\label{eq:PIBD xi_{Am,Sat}}
\end{equation}

7. Obtain the ordinary differential equation (ODE) for the dimensionless barrier corresponding to the current parameters $\delta$, $\alpha$, and $\beta$.

8. Introduce the boundary conditions $\gamma(x_m)$, $\xi(x_m)$, and $\xi_0$ into the ODE, and calculate the absolute value of the dimensionless barrier $\gamma_{Am,Sat}$ at the right endpoint of the ODE.

9. Calculate the temporary saturation voltage at the saturation point:
\begin{equation}
{V_{Sat}} = \left( {{\phi _C} - {\phi _A} - {\gamma _{Am,Sat}}k{T_C}} \right)/e;
\label{eq:PIBD V_{Sat,temp}}
\end{equation}

10. Repeat steps (1-9), incrementally increasing $V$, until $V$ equals $V_{Sat}$ It indicates that the true saturation point has been reached (due to the introduction of the electron recycling effect, the calculated $V_{Sat}$ may vary for different $V$). The operating voltage at this point is the true saturation voltage $V_{Sat}$.

It can be observed that calculating the output characteristics in the order of "incremental voltage" and simultaneously searching for key operating points (saturation points) is an algorithm that can more concisely and smoothly compute the complete "current-voltage" output characteristic curve. For a specific algorithm flowchart, one can refer to Fig.~\ref{fig:fig_PIBD_algor1}.

\subsection{The algorithm of the space-charge limited region}

In the space-charge limited region, the maximum motive $\psi_m$ occurs between the two electrodes. The following steps can be used to find the currents $J_C$ and $J_A$ corresponding to each operating voltage $V$ in the space-charge limited region, and to identify the true critical voltage $V_{Cri}$ as $V$ is continuously increased. This procedure employs a self-consistent iterative algorithm for $\psi_m$, considering the dependence of cathode electron concentration $n$ on the operating voltage, and additionally introduces three parameters, $\delta$, $\alpha$, and $\beta$, for the barrier function that dynamically vary significantly with the operating voltage $V$. The specifics are as follows:

1. Select a voltage $V$ between $V_{Sat}$ and $V_{Cri}$.

2. Choose an initial guess for $\psi_m$ as the maximum motive.

3. Calculate the difference $\gamma(x_m)$ of $\psi_m$ relative to the cathode surface, and the difference $\gamma_{Am}$ of $\psi_m$ relative to the anode surface.

4. Calculate the cathode and anode current densities with the barrier present: $J_C = \frac{n}{n_{eq}}A_RT_C^2\exp\left( - \frac{\phi_C + \psi_{Cm}}{k_BT_C} \right)$ and $J_A = A_RT_A^2\exp\left( - \frac{\psi_m - eV}{k_BT_A} \right)$, and compute $\beta$ using Equation \eqref{eq:PIBD beta}.

5. Obtain the dimensionless barrier ODE corresponding to the current parameters $\delta$, $\alpha$, and $\beta$.

6. Introduce the boundary conditions $\gamma(x_m)$ and $\gamma(d)$ into the ODE, and calculate the dimensionless distance $\xi(d)$ at the right endpoint of the ODE, as well as the difference in dimensionless distance $\xi_{Am}$ between the anode and the maximum barrier.

7. Calculate the dimensionless distance between the two electrodes:
\begin{equation}
\xi_0 = \left( \frac{2\pi m_ee^2}{\epsilon_0^2k^3} \right)^{0.25}\frac{J_C^{0.5}d}{T_C^{0.75}};
\label{eq:PIBD xi_{Am,Sat}}
\end{equation}

8. Repeat steps (2-7) until the error between $\xi_0$ calculated in step 7 and $\xi(d)$ calculated in step 6 is sufficiently small.

9. Repeat steps (1-8), and continuously check the value of $\xi_{Am}$. When $\xi_{Am}$ is infinitely close to 0 (e.g., $< 10^{-3}$), the critical point is reached, and the operating voltage at this point is the critical voltage $V_{Cri}$.

After reaching the critical point, the calculation in the space-charge limited region can be terminated, and the process can proceed to the retrading region. One can refer to Fig.~\ref{fig:fig_PIBD_algor2} in the appendix for the specific algorithm flowchart.

\subsection{The algorithm of the retrading region}

In the retrading region, the maximum barrier $\psi_m$ occurs at the anode surface. At this point, we have ${\gamma_{Am}} = 0$. So at this stage, as $V$ is continuously increased, for each value of $V$, the cathode and anode current densities ${J_C}$ and ${J_A}$ in the presence of the barrier can be directly calculated based on the solutions of the particle rate balance equations. Specifically, ${J_C} = \frac{n}{n_{eq}}A_RT_C^2\exp\left( - \frac{\phi_A + eV}{kT_C} \right)$ and ${J_A} = A_RT_A^2\exp\left( - \frac{\phi_A}{kT_A} \right)$, until $V$ reaches the maximum threshold value we have set.

\bibliography{PETE_PIBD_cite}

\end{document}